# ForceGen: End-to-end *de novo* protein generation based on nonlinear mechanical unfolding responses using a protein language diffusion model

Bo Ni[1], David L. Kaplan[2], Markus J. Buehler[1,3,4]*

[1] Laboratory for Atomistic and Molecular Mechanics (LAMM), Massachusetts Institute of Technology, 77 Massachusetts Ave., Cambridge, MA 02139, USA

[2] Department of Biomedical Engineering, Tufts University, Medford, MA 02155, USA

[3] Center for Computational Science and Engineering, Schwarzman College of Computing, Massachusetts Institute of Technology, 77 Massachusetts Ave., Cambridge, MA 02139, USA

[4]Lead contact

*Correspondence: mbuehler@MIT.EDU

**Abstract:** Through evolution, nature has presented a set of remarkable protein materials, including elastins, silks, keratins and collagens with superior mechanical performances that play crucial roles in mechanobiology. However, going beyond natural designs to discover proteins that meet specified mechanical properties remains challenging. Here we report a generative model that predicts protein designs to meet complex nonlinear mechanical property-design objectives. Our model leverages deep knowledge on protein sequences from a pre-trained protein language model and maps mechanical unfolding responses to create novel proteins. Via full-atom molecular simulations for direct validation, we demonstrate that the designed proteins are novel, and fulfill the targeted mechanical properties, including unfolding energy and mechanical strength, as well as the detailed unfolding force-separation curves. Our model offers rapid pathways to explore the enormous mechanobiological protein sequence space unconstrained by biological synthesis, using mechanical features as target to enable the discovery of protein materials with superior mechanical properties.

**Teaser**: Based on a new force-unfolding dataset derived from full-atomistic molecular modeling, we propose an algorithm that designs novel proteins that meet complex nonlinear mechanical unfolding properties, including detailed unfolding force-separation curves.

## 1. Introduction

Proteins present an elegant yet complex and rich design platform. The various functions and outstanding properties of proteins can be attributed to the folded three-dimensional (3D) structures, encoded by the underling one-dimensional (1D) primary sequences consisting of about 20 naturally occurring amino acids (*1*). Through evolution, nature has demonstrated great success in "designing" proteins as a set of critical building blocks that constitute fundamental functions of all life, and specifically significant biomaterials, ranging from the structural hierarchies in collagens, complex assemblies such as silk, to tissue assemblies such as muscle and skin (*2–5*). In these various tissues and systems, the detailed mechanical signature – often, their response to mechanical pulling – is an essential feature for mechanobiology (*6–9*).

At the same time, there remains vast design space of mechanically optimized proteins yet unexplored by nature given the enormous possibilities of protein sequences (*10*). Hence, inspired by nature, discovering *de novo* proteins may unlock potentially unprecedented properties and functions (*3, 10–16*). However, this enormous design space and costs associated with experimental testing that present great challenges remain in finding effective tools to design *de novo* protein sequences that meet a set of interested functions or properties (*14–19*).

In recent years, the rapid development of deep learning approaches and their applications to proteins have provided fast avenues for protein study and design. For forward problems focused on structure identification, deep learning-based tools such as AlphaFold2 (*20*) and RoseTTAFold(*21*) represents a breakthrough in achieving competitive accuracy with experimental methods in predicting 3D folded structures based on protein sequences at a much reduced cost.(*22*) Built upon these approaches, other protein folding tools (*e.g.*, Omegafold (*23*), RGN2 (*24*), HelixFold-single(*25*) and ESMFold (*26*)) have been exploring the application of large language models. By removing dependence on multiple sequence alignments (MSAs) as the input, improvements in further reducing computational costs and achieving better predictions for orphan and rapidly evolving proteins have been demonstrated (*23–26*).

End-to-end models based on deep learning that predict various structural features (*e.g.*, secondary structure type and content (*27–32*), binding sites (*33*) and surfaces (*34*)) and properties (*e.g.*, solubility (*16*, *35*, *36*), melting temperature (*37*), natural vibrational frequencies (*38*, *39*) and strength(*40*)) for given sequences have also been reported. At the sample time, the inverse design of *de novo* proteins that meet desired structural or property features present a more challenging task. On one hand, facing the enormous sequence space, search algorithms teamed with efficient deep learning-based forward predictors (*29*, *41*, *42*) may still suffer from inefficient exploration and the design accuracy and varieties of the discovered sequences are not easily controlled. On the other hand, recently emergent generative models (*43–48*) provide a direct map from the desired characteristics to potential designs and are becoming a novel paradigm for various materials research and design (*49–54*), including proteins. For example, using an attention-based diffusion model trained on secondary structure data, *de novo* protein sequences can be generated based on secondary structure design objectives (*55*). However, these generative design models often focus on structural level design (such as, secondary structures (*55*) or detailed protein backbone shapes(*56–59*)). In contrast, development of generative models aimed at end-to-end design from property of interest to protein sequence remain rare (*60*).

In this paper, we focus on nanomechanical properties (*61–64*) of proteins. Thanks to the advent of single-molecule technology (*65*) (*e.g.*, atomic force microscopy (*66*, *67*), optical tweezers (*68*, *69*) and magnetic tweezers (*70*, *71*)), the measurement of protein unfolding under an applied mechanical force provides a unique molecular basis for understanding protein deformation (elasticity/plasticity) and fracture (*63*) and can play key roles in affecting some macroscopic mechanical properties of protein-based materials due to the inherent structural hierarchy. For example, via experimental measurements and theoretical analysis, it has been demonstrated that toughness of synthetic protein hydrogels can be correlated to the mechanical unfolding responses of the protein molecules and that mechanically strong folded proteins can result in tough hydrogel designs (*72*). Therefore, generating *de novo* proteins that meet desired mechanical unfolding responses can represent a key molecular level design step in protein-based material designs. Compared with previous protein design cases, this problem presents some unique challenges. First, this is a property-to-sequence end-to-end design task bypassing the structure level, which is expected to be more difficult than previous structure-to-sequence design tasks (*55*, *59*) Second, the available or affordable data on mechanical unfolding responses of known proteins (*73*) are rare when compared to those for protein structures (*74*) or sequences (*75*). Besides mechanical properties, we expect these two challenges are also shared by many other property-to-sequence design tasks in proteins.

To address this problem, in this paper, we combine an attention-based (*76*) diffusion model (*55*) with a pretrained large language model (*26*) for proteins to construct a generative deep learning model that predicts amino acid sequences and 3D protein structures based on mechanical unfolding responses as design objectives. In a singular workflow (**Figure 1**), we start with performing a large series of full-atom molecular dynamics (MD) to simulate the mechanical unfolding process of PDB (*77*) proteins and recording the force responses (**Figure 1**a). Then, we construct a protein language diffusion model (pLDM) by translating the protein sequences into a word probability latent space using a pretrained protein language model and training a diffusion model to learn the map between sequence representations and the force-separation responses (**Figure 1**b). At deployment, the trained pLDM predicts sequence candidates based on the given unfolding force conditions and the integrated folding algorithm (*i.e.*, OmageFold) (23) determines the 3D structures of the resulting sequences. For validation, we compare the designed sequences with known proteins to analyze novelty (**Figure 1**c) and to test the designed proteins using MD to compare the mechanical properties and unfolding responses with input conditions (**Figure 1**d). Through well-controlled comparisons, we demonstrate that our pLDM outperforms the vanilla diffusion model with or without an iterative design scheme. Built upon the property-to-sequence generation capability of our model and the broad potential of protein materials in achieving superior mechanical properties, as well as other interesting properties (78–81) (e.g., optical (80, 81), electronic (78), energy storage (79), etc.) we expect our end-to-end design model can be useful in numerous biological and engineering applications for the property-targeted generative design of various protein material systems.

## 2. Results and Discussion

### Full-atom modeling of unfolding proteins by force

Inspired by single-molecule force-spectroscopy(66), we simulate the unfolding process of protein chains under mechanical force to understand their mechanical properties at the molecular level. As shown in **Figure 2a**, we start with PDB proteins with experimentally measured 3D structures. Using full-atom MD with the CHARMM force field(82) and a generalized Born implicit solvent model (83), we first relax the protein molecule at body temperature (*i.e.*, 310K) to reach equilibrium conformation. Then, we stretch the protein chain of $N$ amino acids along the direction connecting the two chain ends (*i.e.*, $a$ and $b$) by fixing one end and steering the other with a spring (*i.e.*, the segment between $b$ and $c$ in **Figure 2a**) of a force constant $k = 0.5\,\mathrm{kcal/(mol\ Å^2)}$ at a constant velocity $v = 0.1\,\mathrm{Å/ps}$. The pulling force, $F_p$, is recorded every 0.2 ps until the distance between two pulling ends, $L_{ac}$, reaches the contour length, $L_{con}$, of the protein chain, where we assume the average length of each amino acid is 3.6 Å (84) and $L_{con} = N \times 3.6$ Å. Further details on the MD simulations can be found in the **Materials and Methods** section. Movies of the unfolding trajectory of some selected PDB protein examples can be found in in the **Supplementary Materials**.

In **Figure 2b**, we smooth the raw force response (the red curve) to get rid of high frequency fluctuations and get the unfolding response, $F_p(L_{ac})$, of the protein chain (the blue curve), from which we can identify the toughness and strength of the protein molecule using the unfolding energy $T$ and the maximal value of force $F_{max}$ defined as the following.

$$T = \int^{L_{con}} F_p dL_{ac} \tag{1}$$

$$F_{max} = \max_{L_{ac} \leq L_{con}} \{F_p\} \tag{2}$$

To curate a dataset based on naturally existing proteins, we use the Biomolecule Stretching Database (BSDB) (73) as a guidance and select 7,026 PDB proteins that have no gaps in their experimentally determined structures and consist of no more than 128 amino acids. Then, we collect their structures directly from the Protein Data Bank (77) and apply the protocol above to test their mechanical unfolding responses. An overview of the distributions of the unfolding responses and mechanical properties are shown in **Figure 2c-d**. Specifically, in **Figure 2d**, the unfolding energy or toughness shows a bimodal distribution while the strength presents a unimodal one; in **Figure 2c**, one can observe that there exist various unfolding responses among proteins. For example, the maximal force may appear as the peak in the middle of unfolding process or near the end when reaching the contour length, which may indicate very different deformation mechanisms. An in-depth study of these mechanical properties and their dependence on the structural features and sequences for a large number of proteins deserves a separate study and is left for future work. Further insight can also be obtained from the numerous experimental studies using AFM-based force generative outcomes during unfolding states of proteins. Here, we focus on applying this newly collected data to develop protein generation models. To efficiently label the pulling force responses during the full unfolding process, we introduce a pulling force vector $\vec{F}_p$ (green triangles in **Figure 2b**) to represent the full response as the following

$$\vec{F}_p = \{F_p(L_{ac}^i): i = 0,1,2,\dots,N\}\,, \quad (3)$$

where $N$ is the sequence length of the protein chain and we sample the pulling force when the distance between the pulling ends reaches $L_{ac}^i = i \times L_{con}/N$. For $L_{ac}^i$ that is smaller than the value of the initial equilibrium conformation, we simply define the force values as zeros. Such a vector representation can adjust to the protein sequence length automatically, that is, longer/shorter proteins with potentially more/fewer unfolding details have more/fewer sampling points evenly distributed. Next, we develop DL models to generate protein sequences that meet the given mechanically unfolding responses represented in terms of the pulling force vector $\vec{F}_p$.

**Protein language diffusion model and inverse design for mechanical signatures**

To solve the conditioned protein design task, here we develop a protein language diffusion model (pLDM) by combining a pretrained protein language model (pLM) (*26*) and an attention-based diffusion model(*55*). **Figure 3a** depicts an overview of the model developed in the present work. The pLM (on the right of **Figure 3a**) is pretrained on large amounts of protein sequences data (*85*) to form internal representations that better understand not only sequences but also structures and properties of proteins.(*32*) We leverage this knowledge by applying pLMs to translate protein sequences from the tokenized sequence space into the word probability latent space. Then, we train a diffusion model developed in the previous work to operate in this probability latent space. The diffusion model is built upon a one-dimensional U-Net architecture with attention mechanisms (**Figure 3b**). At deployment, starting with the given condition (on the left of **Figure 3a**) and random signal seed, the diffusion model predicts and removes the noise at each step and produces meaningful sequence probability tensors, which are then translated back into protein sequence using the fixed pLM. There are multiple choices for the pretrained pLM and larger pLMs require higher computing resource and cost.(*26, 32, 86*) For computational efficiency, in the following, we focus on the results that adopt an medium-sized pretrained model with 150M parameters from the ESM-2 series (we find that this yields improved performances as is shown in the discussion at the next subsection) (*87*).

Once the model has been trained, we demonstrate the performance of the developed pLDM by testing it with various mechanical unfolding responses, including those that come from naturally existing proteins and those that are *de novo*. The generated sequences are folded into 3D structures using OmegaFold(*23*) and then undergo the same mechanical unfolding tests using full-atom MD simulations. With protein BLAST(*88*) test and comparing pulling force responses with the input, we exam the novelty of the generated sequences and the accuracy of the protein design.

For protein design with the mechanical unfolding responses that correspond to naturally existing proteins, we test the model with the pulling force records of PDB proteins in the test set, with which the model has not been trained. **Figure 4** shows some examples of the designed proteins and their mechanical unfolding responses. In terms of the design target, the conditioned pulling force paths (red curves) in **Figure 4a-f** represent a variety of different patterns, including simple ones that show the pulling force nearly keeps increasing during the unfolding process (**Figure 4**d), the examples that show local peaks in early stages of unfolding during an overall increasing trend of the pulling force (**Figure 4**a, b and f), the ones that reach an oscillating plateau and then increases (**Figure 4**e) and those that achieve high peak in the initial stage of unfolding (**Figure 4**c). Despite this complexity of protein unfolding scenarios, and the oscillating nature of pulling force, the proteins generated by our model demonstrate pulling force responses (blue curves) that in general closely follow the design objectives. We use multiple metrics, including $R^2$ and relative $L_2$ error (see **Materials and Methods** section for details), to measure the accuracy of design in meeting the overall trend as well as quantitative values of the mechanical responses. Corresponding to the various patterns of pulling force responses, the generated proteins also show a variety of internal structures. For example, a relatively simple combination of alpha-helix segment and random coils without complex entanglement or close spatial packing in **Figure 4d** produces an unfolding process with consistently - increasing pulling forces with few oscillations. An entangled mix of alpha-helix and beta-sheet in **Figure 4e** yields an unfolding pulling force path with a plateau region of strong oscillations. Unfolding of the beta-sheet in **Figure 4c** can be related to the high local peak in pulling force history. In **Figure 4a**, **b** and **f**, alpha-helix segments of different lengths and spatial arrangements can produce similar trends but different quantitative pulling force records. More details on the mechanical unfolding process of some of these cases, including cases **a**, **c** and **e**, can be found in the MD trajectory movies in the **Supplementary Materials**.

On the novelty of these generated proteins, we apply basic local alignment search tool (BLAST) analysis (*88*) to the predicted amino acid sequences to access whether, and to what extent, they represent novel sequences or closely related forms of known proteins. **Table 1** shows the results of the BLAST analysis for the various cases listed in **Figure 4**. We find that even though the input design targets are from existing PDB proteins, many of the generated protein sequences (cases shown in **Figure 4a-d** and **f**) do not match any sequences in the database of known proteins with standard BLSAT analysis (i.e., returning "no significant similarity found" in protein BLAST test) and are *de novo* ones. The model can also produce sequences (*e.g.*, case **e** in **Figure 4**) that show some similarity to the existing proteins. However, the most similar example found (*i.e.*, 8CH0) is not included in the training and testing set. While the model is only trained on a very small portion of PDB proteins, with the pulling force corresponding to existing PDB proteins as an input, we expect the possibility of the model "rediscovering" sequences that show some similarities to the known proteins. Further measures may be utilized to boost the novelty of design for such cases, including multi-shot design and selecting the best one based on BLAST results. In current work, we focus on understanding the performance of the current model.

Besides examining individual cases, we also show the distributions of design accuracy and novelty for a larger number of testing cases. **Figure 5** demonstrates the results of 187 generated proteins based on various pulling force conditions from the standalone test set. On the mechanical unfolding responses, the $R^2$ and relative $L_2$ error between the measured pulling force vectors of the generated proteins and the input conditions among cases show unimodal distributions with the median values of 0.56 and 0.36 (**Figure 5a** and **b**), respectively. The distributions indicate that for many of the cases, the designed proteins follow the input conditions in term of the trend and value reasonably well during for the whole mechanical unfolding process. However, as demonstrated in individual cases (**Figure 4**), it remains challenging for designed proteins to precisely follow the input pulling force values at each unfolding step. This can also be seen by comparing components of pulling forces of all cases together in **Figure 5c**. While conditioned input values of pulling force components and the measured ones based on the generated proteins in general share the same trend (that is, the distribution centers at the $y=x$ dash line as the visual guide in **Figure 5c**), the finite width of the distribution cloud deviating away from the ideal case indicates that for individual components, there could be considerable mismatches.

The limited component-wise accuracy demonstrates the difficulty and challenge of designing proteins based on detailed mechanical unfolding responses, even with the current model. At the same time, the proteins generated by our model still show reasonable agreement between the achieved and the conditioned mechanical properties, including toughness (**Figure 5d**) and strength (**Figure 5e**). Strength defined as the maximal of the pulling force shows a $R^2$ value of 0.41 (**Figure 5e**), slightly smaller than that of the pulling force components (0.54 as listed in **Figure 5c**). At the same time, a $R^2$ value of 0.93, much higher than that of pulling force components (**Figure 5c**), was observed for toughness (**Figure 5d**), which is defined as the unfolding energy over the whole unfolding process (Equation (1)). This difference in $R^2$ values indicates that when the entire unfolding process is considered, the component-wise error tends to cancel each other and the designed proteins follows the input conditions in terms of toughness more sensitively. On the novelty of the designed proteins, **Figure 5f** shows a bimodal distribution of the highest percent identity found via protein BLAST analysis for all the generated sequences. The highest peak (on the left in **Figure 5f**) corresponds to the cases where the generated proteins have little similarity to the existing/known ones and totally *de novo*. There also exists the other weaker peak on the right for cases in which the generated proteins are similar to the known ones. The bimodal distribution echoes the result of individual cases listed in **Table 1** and the relative height of the two peaks indicates our model has a stronger tendency in generating *de novo* sequence designs.

To further boost the novelty of designed sequences, different measures could be taken, including how the model is applied and how the input conditions are constructed. Here, we discuss one possibility of using *de novo* mechanical unfolding responses, *i.e.*, pulling force vectors that do not correspond to existing proteins, as the input. There is still a lack of complete rules on how to identify or construct physically achievable pulling force vectors for all possible amino acid sequences. To explore the *de novo* mechanical unfolding responses, here we start with existing pulling force vectors and mute them using a mixing scheme. As shown in **Figure 6a**, two pulling force vectors from PDB proteins 2AAN and 1KN7 were chosen and show different patterns during the unfolding process, one undergoing an oscillating plateau in the early stages while the other keeps increasing. These differences in pulling force can be attributed to the different internal protein structures and sequence length. 2AAN shows a high beta-sheet content in a closely packed conformation. In comparison, 1KN7 has a mix of alpha-helix and unstructured coils with a more open spatial arrangement. We mix the pulling force vectors of each of the proteins with respect to the same normalized pulling end gap at different ratios (*i.e*., 1/3 to 2/3 for Mix 1 and 2/3 to 1/3 for Mix 2) and then use these as the *de novo* mechanical unfolding response to generate proteins with our models. Also, in our model, the length of the generated sequence (*i.e.*, $N$) can be controlled through the length of the input pulling force vector (*i.e.*, N+1). Here, we intentionally choose $N = 99$, which is different from that of both base proteins when constructing the *de novo* input pulling force vectors. The results of the proteins designed with these *de novo* pulling force vectors are shown **Figure 6b** and **c**. Similar to previous designs listed in **Figure 4**, the mechanical unfolding responses of the new designs follow the input conditions, even though this time they come from a mix of proteins of different structures and mechanical properties. Also, the designed proteins adopt internal structures of a mix of alpha-helix with different compactness, which show little similarity to those of the base proteins. Another set of examples is shown in **Figure 6d-f** with base proteins of different internal structures. Again, the newly designed proteins fulfill the targeted *de novo* mechanical unfolding responses reasonably well.

With the obtained *de novo* pulling force responses as the input, it is interesting to investigate the novelty of the generated proteins. As shown in **Table 2**, all of the designs have no significant similarity among known protein sequences according to protein BLAST analysis. The case study shown here demonstrates that our generative model can also be used as an effective tool to explore the unknown possibilities in property space of mechanical unfolding responses by constructing *de novo* input conditions. In return, it is more probable to discover *de novo* designs of new proteins.

Besides using the mixing scheme demonstrated above, other ways can be used to create or discover novel mechanical unfolding responses as the design input. One possibility is to train a generative model (*e.g.*, a diffusion model) on the pulling force data curated here. At deployment, mechanical unfolding responses can be generated with or without input conditions. Because the possible pulling force vectors should in principle be determined by the sequence-structure-property relationship of proteins, in-depth investigation on this process deserves a separate study, where the generation model developed here can serve as an effective tool in navigating the complex design space for proteins.

### Protein language diffusion model benefits small-data generation tasks

Combining the above results of protein designs based existing or *de novo* mechanical unfolding responses, we have demonstrated that the pLDM developed here can achieve reasonably good design accuracy for both detailed unfolding pulling force history and overall mechanical properties. At the same time, many of the generated sequences are totally *de novo*, having no significant similarity among any existing or known proteins. This is quite surprising considering the fact that: i) the design task is challenging, and ii) the training data used is relatively small comparing with previous work. Specifically, the design task here is to map detailed pulling force responses directly to protein sequences, bypassing predicting the 3D structural transition of protein chains during the unfolding process. Based on MD simulations and experimental studies, it is clear that the unfolding process of protein structures under mechanical forces corresponds to complex uncoiling of the backbone and breaking of hydrogen bonds within the 3D hierarchical structures of proteins. Theoretically, we can expect that this design task could be more challenging than designing protein sequences based on structural conditions (*e.g.*, secondary structure types). At the same time, for the pLDM discussed above which designs sequences with a length, $N$, smaller than or equal to 126, it was trained with a data set of 6,863 proteins, mainly limited by the high computational cost to curate the data on mechanical properties. The size of the dataset is quite small compared with the PDB proteins of known structures (*i.e.*, about 13,644 for single-chain protein with $N \leq 126$ ), and even further dwarfed by the number of all possible protein sequences (*i.e.*, more than $20^{126}$).

Given such a challenging protein design problem with a limited small set of labelled data, we discuss the strength of the current protein language diffusion model (pLDM) by comparing it with the protein diffusion models (pDM) developed previously.(*55*) The structure of the adjusted pDM is summarized in **Figure S1a**. The main difference between pLDM (**Figure 2b**) and pDM is about the space where the diffusion/de-noising processes occur during the training/inferring. While the pDM operates in the tokenized protein sequence space, pLDM performs in the word probability latent space formed by pre-training the pLM on the available protein sequence dataset (*i.e.*, UniRef50 dataset for ESM-2 pLM adopted here), which is much larger than the datasets on protein structures or mechanical properties. For various downstream prediction tasks, the adoption of the pretrained pLM as a base model has proven to be beneficial (*32*). For example, the ESM-2 model adopted here has been used for protein folding prediction in ESMFold (*26, 86*) and achieved comparable accuracy with the folding tool using MSA as the input (*e.g*., AlphaFold2) but at a lower computational cost. At the same time, for inverse generation, the integration of a large language model and diffusion model has proven to be beneficial in conditioned generation tasks for images(*47, 89*) and texts (*90, 91*) (*e.g.,* Diffusion-LM can achieve fine-grained controls on syntactic structure for text generation (*91*)). However, the applications of pretrained pLM in the generative tasks of protein design remain rare(*92*) and the potential benefits are to be explored. Here, the mechanical unfolding response of the conditioned protein design task and our results provide a concrete example to study this aspect of the process.

To compare with pLDM, two alternative models based on pDMs were constructed. The first one, AM1, uses one pDM and designs the protein sequence for a given pulling force vector in one shot, similar to the pLDM developed above. The second model consist of two pDMs as the protein designer and protein predictor separately. The designer is the same as the first pDM model while the predictor is trained to predict pulling force vector based on given protein sequences. At deployment, the protein designer iteratively generates sequence candidates and the protein predictor then evaluates them to pick the most accurate one among five attempts, forming an on-the-fly iterative design scheme (**Figure S1b**). Both models are trained on the same dataset discussed earlier and tested with the same generation task based on existing mechanical unfolding responses from the test set. A summary of the performance of the two pDM-based models together with those of the pLDM is listed in **Table 3**. We use various metrics, $R^2$ and relative errors, to evaluate the design accuracy of the models on the whole test set, considering not only the overall mechanical properties (*i.e.,* toughness and strength) but also the detailed mechanical unfolding responses in terms of the pulling force vectors. For most of the comparisons (9 out the 11 rows), pLDM achieves the best performance (*e.g.*, the highest $R^2$ or the lowest error). This result demonstrates that, by integrating the pretrained protein language model, pLDM achieves enhancement in producing more accurate designs even when being trained only on a relatively small dataset. This advantage of pLDM with respect to pDM can be quite helpful for other property-to-sequence design tasks for proteins, given that usually data on protein mechanical or other properties are more limited or costly to collect.

## 3. Conclusions

Generating novel proteins based on their mechanical unfolding responses presents unique challenges in property-targeted protein design. For rational design strategies, it is hard to grasp the complex relationships between sequences, structures and properties. For data-driven methods, the labelled data on the mechanical properties are often costly to collect and limited in number, especially given the enormous possibilities in protein sequence space.

Here, we have developed a pLDM as an effective tool to tackle these challenges and generate novel proteins that meet the mechanical properties design objectives in an end-to-end manner. The pLDM developed combines pretrained pLM and diffusion model as key components and leverages the strength of both. The pLM part is pretrained on the abundant protein sequence data and thus provides an effective representation of protein sequences in its latent space. The diffusion model part only operates in this latent space and learns the map between detailed pulling force responses and the sequence representation using only a relatively small set of data curated by performing full-atom simulations. By examining the unfolding details of individual designs and the statistics of the mechanical properties of many cases, we demonstrate that the proteins designed by our model meet the targeted overall mechanical properties, including toughness and strength, as well as the detailed unfolding force vectors with reasonably good accuracy. Moreover, the sequences generated are mostly *de novo*, sharing very limited similarity with existing/known proteins. Given the mechanical unfolding responses from known PDB proteins as the design input, our model still shows a strong tendency in discovering *de novo* proteins as alternatives. Constructing *de novo* unfolding responses as the input via a mixing scheme further boosts the probability of generating *de novo* designs. Finally, through controlled comparisons, we show that the pLDM outperforms the vanilla protein diffusion model with or without an iterative design scheme in achieving better design accuracy, thus clearly demonstrating the benefits of combining pretrained pLM and diffusion model in the pLDM developed here. A short summary of these key aspects about the pLDM is listed in **Table 4**.

The pLDM we developed here offers a unique and powerful means to explore the enormous protein sequence spaces with molecular mechanical properties as guidance, to meet specific mechanobiology properties. Applying our model, future studies can start with a systematic study on the relationships between sequence-structure-mechanical properties in proteins. For example, with our model one can gradually change the pattern of pulling force among various known cases (*e.g.*, **Figure 2c**), to generate corresponding protein candidates and to study their structural pattern transitions and sequence mutations. At the same time, as the designed *de novo* proteins increases in number, their sequences and mechanical unfolding responses can be used as new data to gradually increases the protein dataset on mechanical properties and our model can be further trained on this growing set. With the more powerful model, one can further study some challenging topics, such as designing proteins with the optimal mechanical properties or even their combination in various engineering and biological applications.

While we have developed the pLDM that takes the mechanical unfolding responses as the design conditions here, we expect similar pLDM frameworks can be generalized for other property-to-sequence design tasks in proteins. The enhancement brought by merging pretrained pLMs can be inspiring for other design tasks, especially where only small data sets on the property of interest are available or affordable at the beginning. At the same time, going beyond only one type of condition as the design target, our pLDM can also be generalized for design tasks under multiple objectives, given the flexibility of the diffusion model in incorporating these conditions (**Figure 3b**). Combining the previous work using a protein diffusion model, one example can be taking both secondary structure and unfolding forces as the design target. Also, during the generation process, techniques like inpainting through selective masking or biasing certain amino acids(*93*) are straightforward to implement. Combining these under the pLDM framework, we envision a comprehensive generative model that moves towards designing proteins at all levels, including sequence, structure and properties in harmony.

## 4. Materials and Methods

### Protein mechanical unfolding simulations by molecular dynamics

We use NAMD to perform full-atom molecular dynamics simulations. The interaction between protein atoms is described by the CHARMM force filed (*82*). We adopt a generalized Born implicit solvent model (*83*) for the effect of solvent on proteins. Compared with simulations with an explicit solvent model, our setup balances the accuracy and the computational costs. We develop a parallel workflow to simulate mechanical unfolding process of about 7,026 proteins of various sequence length.

### Dataset

We curate the dataset based the MD results. Key information for each protein case includes, PDB ID, protein sequence, sequence length, pulling force vector, strength and toughness. See **Figure 2** for details on their distributions. We use 85% of the dataset for training and keep 15% for testing.

### Design of the neural network architectures and training

The protein language diffusion model developed here consists of a pretrained protein language model (pLM) and a diffusion model. Only the latter is trainable.

For the pretrained pLM, we use one variant with 150M parameters from the ESM-2 series (*26, 87*). During training, we first propagate a mini-batch of $B$ tokenized sequences with a length smaller than or equal to N in terms of a Bx1xN tensor to the last hidden layer of the pretrained pLM to get the logits, then normalized them into a BxMxN tenor, where the M components in the 2nd dimension represent the probability of that position being each of the M words in the pLM model. The diffused model is based the previous work and only operates in this word probability space introduced by the pLM. At the deployment, the output of the diffusion model will be translated back to the tokenized sequences by picking the word with the highest probability.

**Protein folding**

We adopt OmegaFold (*23*) for rapid prediction of protein structures from the sequence. OmegaFold offers a rapid alternative as it does not require Multiple Sequence Alignment (MSA), yet produces results of similar accuracy as AlphaFold2(*20*) and trRosetta(*94*) (and similar, related state of the art methods).

**Design accuracy evaluation**

We use various metrics to compare the measured mechanical unfolding responses and mechanical properties with the input design conditions for individual designs as well as predictions for the whole test set.

For vectors, including the unfolding pulling force vector for one protein and toughness or strength for proteins in the test set, the $R^2$ and relative $L_2$ error defined as the following,

$$R^2\,[\vec{x},\vec{y}] = 1 - \frac{\sum_i (x_i - y_i)^2}{\sum_i (x_i - \bar{x})^2} \tag{4}$$

$$L_2^{rela}[\vec{x},\vec{y}] = \frac{\|\vec{x}-\vec{y}\|}{\|\vec{x}\|} = \frac{\sqrt{\sum_i (x_i - y_i)^2}}{\sqrt{\sum_i (x_i)^2}} \tag{5}$$

where $\vec{x}$ is the ground truth or input vector and $\vec{y}$ is the measured one from the predictions, $x_i$ and $y_i$ are their components and $\bar{x}$ is the mean of the components $x_i$.

For scalars, including the toughness and strength for one protein, the relative $L_1$ error is defined as the following,

$$L_1^{rela}[x,y] = \frac{|y-x|}{|x|} \tag{6}$$

where $x$ is the ground truth or input value and $y$ is the measured value based on the prediction.

**BLAST analysis**

The basic local alignment search tool (BLAST) analysis (*88*) for the various cases is conducted using the blastp (protein-protein BLAST) algorithm, and the non-redundant protein sequences (nr) database.

**Visualization**

We use Visual Molecular Dynamics (VMD) (*95*) for visualization of the protein structures.

**Software versions and hardware**

We use Python 3.9.16, PyTorch 1.12.1+cu13(*96*) with CUDA (CUDA version 12.0), and a NVIDIA Tesla V100 with 32 GB VRAM for training and inference.

**Acknowledgments:** We acknowledge support from USDA (2021-69012-35978), DOE-SERDP (WP22-S1-3475), ARO (79058LSCSB, W911NF-22-2-0213 and W911NF2120130) as well as the MIT-IBM Watson AI Lab and

MIT's Generative AI Initiative. Additional support from NIH (U01EB014976 and R01AR077793) and ONR (N00014-19-1-2375 and N00014-20-1-2189) is acknowledged.

## Conflict of interest

The author declares no conflict of interest.

## Data availability

Data and codes, as well as trained weights, are either available on GitHub at https://github.com/lamm-mit/ProteinMechanicsDiffusionDesign or will be provided by the corresponding author based on reasonable request.

**Author contributions:** MJB conceived the study. BN curated the dataset, developed and trained the neural network and performed associated data analysis and prepared the first draft. MJB and DLK supported the analysis and wrote the paper with BN.

## Supplementary materials

Additional figures, PDB files, and other materials are provided as **Supplementary Materials**.

- CSV files of the curated dataset on protein sequences and mechanical unfolding responses used for training and validation cases.
- ZIP files with **PDB proteins structures** generated by the pLDM with some representative mechanical unfolding responses (**Figure 4** and **6**) are included as **Supplementary Material**.
- Movies for the unfolding process of selected PDB proteins from the curated dataset: https://www.dropbox.com/scl/fi/33tnpd6u2xwermlvj22y9/SI_3_unfolding_movies_from_dataset.zip?rlkey=qno7rcitcdree8t9cj8wzg9sf&dl=0
- Movies of the unfolding process of selected *de novo* proteins designed by the pLDM: https://www.dropbox.com/scl/fi/gjpnfsltmd9eh61mhxud8/SI_4_unfolding_movies_from_design.zip?rlkey=rvvfjfz4o2itv0pps3akwik9y&dl=0

## References

1. Protein Structure and Function - Gregory A. Petsko, Dagmar Ringe - Google Books. https://books.google.com/books?hl=en&lr=&id=bCI5u_19N_oC&oi=fnd&pg=PR5&ots=aGbgx6l-Qc&sig=Ciae9YbL1Bxi2mpAhonAu93TVPQ#v=onepage&q&f=false.

2. D. López Barreiro, J. Yeo, A. Tarakanova, F. J. Martin-Martinez, M. J. Buehler, D. López, J. Yeo, A. Tarakanova, F. J. Martin-Martinez, M. J. Buehler, Multiscale Modeling of Silk and Silk-Based Biomaterials—A Review. *Macromol Biosci* **19**, 1800253 (2019).

3. G. Gronau, S. T. Krishnaji, M. E. Kinahan, T. Giesa, J. Y. Wong, D. L. Kaplan, M. J. Buehler, A review of combined experimental and computational procedures for assessing biopolymer structure–process–property relationships. *Biomaterials* **33**, 8240–8255 (2012).

4. C. Vepari, D. L. Kaplan, Silk as a biomaterial. *Prog Polym Sci* **32**, 991–1007 (2007).

5. S. Ling, D. L. Kaplan, M. J. Buehler, Nanofibrils in nature and materials engineering. *Nature Reviews Materials 2018 3:4* **3**, 1–15 (2018).

6. K. A. Jansen, D. M. Donato, H. E. Balcioglu, T. Schmidt, E. H. J. Danen, G. H. Koenderink, A guide to mechanobiology: Where biology and physics meet. *Biochimica et Biophysica Acta (BBA) - Molecular Cell Research* **1853**, 3043–3052 (2015).

7. A. E. M. Beedle, S. Garcia-Manyes, The role of single-protein elasticity in mechanobiology. *Nature Reviews Materials 2022 8:1* **8**, 10–24 (2022).

8. J. L. Balestrini, J. K. Skorinko, A. Hera, G. R. Gaudette, K. L. Billiar, J. L. Balestrini, J. K. Skorinko, · A Hera, · G R Gaudette, · K L Billiar, Applying controlled non-uniform deformation for in vitro studies of cell mechanobiology. *Biomechanics and Modeling in Mechanobiology 2010 9:3* **9**, 329–344 (2010).

9. Z. Qin, M. J. Buehler, L. Kreplak, A multi-scale approach to understand the mechanobiology of intermediate filaments. *J Biomech* **43** (2010).

10. P. S. Huang, S. E. Boyken, D. Baker, The coming of age of de novo protein design. *Nature 2016 537:7620* **537**, 320–327 (2016).

11. U. G. K. Wegst, H. Bai, E. Saiz, A. P. Tomsia, R. O. Ritchie, Bioinspired structural materials. *Nature Materials 2014 14:1* **14**, 23–36 (2014).

12. G. X. Gu, M. Takaffoli, M. J. Buehler, G. X. Gu, M. Takaffoli, M. J. Buehler, Hierarchically Enhanced Impact Resistance of Bioinspired Composites. *Advanced Materials* **29**, 1700060 (2017).

13. F. Barthelat, Z. Yin, M. J. Buehler, Structure and mechanics of interfaces in biological materials. *Nature Reviews Materials 2016 1:4* **1**, 1–16 (2016).

14. W. Huang, A. Tarakanova, N. Dinjaski, Q. Wang, X. Xia, Y. Chen, J. Y. Wong, M. J. Buehler, D. L. Kaplan, Design of Multistimuli Responsive Hydrogels Using Integrated Modeling and Genetically Engineered Silk–Elastin-Like Proteins. *Adv Funct Mater* **26**, 4113–4123 (2016).

15. S. T. Krishnaji, G. Bratzel, M. E. Kinahan, J. A. Kluge, C. Staii, J. Y. Wong, M. J. Buehler, D. L. Kaplan, Sequence–Structure–Property Relationships of Recombinant Spider Silk Proteins: Integration of Biopolymer Design, Processing, and Modeling. *Adv Funct Mater* **23**, 241–253 (2013).

16. M. J. Buehler, Generative pretrained autoregressive transformer graph neural network applied to the analysis and discovery of novel proteins. *J Appl Phys* **134**, 84902 (2023).

17. A. Paladino, F. Marchetti, S. Rinaldi, G. Colombo, Protein design: from computer models to artificial intelligence. *Wiley Interdiscip Rev Comput Mol Sci* **7**, e1318 (2017).

18. Z. Qin, L. Wu, H. Sun, S. Huo, T. Ma, E. Lim, P. Y. Chen, B. Marelli, M. J. Buehler, Artificial intelligence method to design and fold alpha-helical structural proteins from the primary amino acid sequence. *Extreme Mech Lett* **36**, 100652 (2020).

19. J. Wang, H. Cao, J. Z. H. Zhang, Y. Qi, Computational Protein Design with Deep Learning Neural Networks. *Scientific Reports 2018 8:1* **8**, 1–9 (2018).

20. J. Jumper, R. Evans, A. Pritzel, T. Green, M. Figurnov, O. Ronneberger, K. Tunyasuvunakool, R. Bates, A. Žídek, A. Potapenko, A. Bridgland, C. Meyer, S. A. A. Kohl, A. J. Ballard, A. Cowie, B. Romera-Paredes, S. Nikolov, R. Jain, J. Adler, T. Back, S. Petersen, D. Reiman, E. Clancy, M. Zielinski, M. Steinegger, M. Pacholska, T. Berghammer, S. Bodenstein, D. Silver, O. Vinyals, A. W. Senior, K. Kavukcuoglu, P. Kohli, D. Hassabis, Highly accurate protein structure prediction with AlphaFold. *Nature 2021 596:7873* **596**, 583–589 (2021).

21. M. Baek, F. DiMaio, I. Anishchenko, J. Dauparas, S. Ovchinnikov, G. R. Lee, J. Wang, Q. Cong, L. N. Kinch, R. Dustin Schaeffer, C. Millán, H. Park, C. Adams, C. R. Glassman, A. DeGiovanni, J. H. Pereira, A. V. Rodrigues, A. A. Van Dijk, A. C. Ebrecht, D. J. Opperman, T. Sagmeister, C. Buhlheller, T. Pavkov-Keller, M. K. Rathinaswamy, U. Dalwadi, C. K. Yip, J. E. Burke, K. Christopher Garcia, N. V. Grishin, P. D. Adams, R. J. Read, D. Baker, Accurate prediction of protein structures and interactions using a three-track neural network. *Science (1979)* **373**, 871–876 (2021).

22. M. Varadi, S. Anyango, M. Deshpande, S. Nair, C. Natassia, G. Yordanova, D. Yuan, O. Stroe, G. Wood, A. Laydon, A. Zídek, T. Green, K. Tunyasuvunakool, S. Petersen, J. Jumper, E. Clancy, R. Green, A. Vora, M. Lutfi, M. Figurnov, A. Cowie, N. Hobbs, P. Kohli, G. Kleywegt, E. Birney, D. Hassabis, S. Velankar, AlphaFold Protein Structure Database: massively expanding the structural coverage of protein-sequence space with high-accuracy models. *Nucleic Acids Res* **50**, D439–D444 (2022).

23. R. Wu, F. Ding, R. Wang, R. Shen, X. Zhang, S. Luo, C. Su, Z. Wu, Q. Xie, B. Berger, J. Ma, J. Peng, High-resolution de novo structure prediction from primary sequence. *bioRxiv*, 2022.07.21.500999 (2022).

24. R. Chowdhury, N. Bouatta, S. Biswas, C. Rochereau, G. M. Church, P. K. Sorger, M. Alquraishi, Single-sequence protein structure prediction using language models from deep learning. *bioRxiv* **2021-November**, 2021.08.02.454840 (2021).

25. X. Fang, F. Wang, L. Liu, J. He, D. Lin, Y. Xiang, X. Zhang, H. Wu, H. Li, L. Song, HelixFold-Single: MSA-free Protein Structure Prediction by Using Protein Language Model as an Alternative. (2022).

26. Z. Lin, H. Akin, R. Rao, B. Hie, Z. Zhu, W. Lu, N. Smetanin, R. Verkuil, O. Kabeli, Y. Shmueli, A. dos Santos Costa, M. Fazel-Zarandi, T. Sercu, S. Candido, A. Rives, Evolutionary-scale prediction of atomic-level protein structure with a language model. *Science (1979)* **379**, 1123–1130 (2023).

27. M. H. Høie, E. N. Kiehl, B. Petersen, M. Nielsen, O. Winther, H. Nielsen, J. Hallgren, P. Marcatili, NetSurfP-3.0: accurate and fast prediction of protein structural features by protein language models and deep learning. *Nucleic Acids Res* **50**, W510–W515 (2022).

28. B. Zhang, J. Li, Q. Lü, Prediction of 8-state protein secondary structures by a novel deep learning architecture. *BMC Bioinformatics* **19**, 1–13 (2018).

29. C. H. Yu, W. Chen, Y. H. Chiang, K. Guo, Z. Martin Moldes, D. L. Kaplan, M. J. Buehler, End-to-End Deep Learning Model to Predict and Design Secondary Structure Content of Structural Proteins. *ACS Biomater Sci Eng* **8**, 1156–1165 (2022).

30. G. Pollastri, A. McLysaght, Porter: a new, accurate server for protein secondary structure prediction. *Bioinformatics* **21**, 1719–1720 (2005).

31. C. Mirabello, G. Pollastri, Porter, PaleAle 4.0: high-accuracy prediction of protein secondary structure and relative solvent accessibility. *Bioinformatics* **29**, 2056–2058 (2013).

32. A. Elnaggar, M. Heinzinger, C. Dallago, G. Rehawi, Y. Wang, L. Jones, T. Gibbs, T. Feher, C. Angerer, M. Steinegger, D. Bhowmik, B. Rost, ProtTrans: Toward Understanding the Language of Life Through Self-Supervised Learning. *IEEE Trans Pattern Anal Mach Intell* **44**, 7112–7127 (2022).

33. J. Tubiana, D. Schneidman-Duhovny, H. J. Wolfson, ScanNet: an interpretable geometric deep learning model for structure-based protein binding site prediction. *Nature Methods 2022 19:6* **19**, 730–739 (2022).

34. F. Sverrisson, J. Feydy, B. E. Correia, M. M. Bronstein, Fast end-to-end learning on protein surfaces. *bioRxiv*, 2020.12.28.424589 (2020).

35. V. Thumuluri, H. M. Martiny, J. J. Almagro Armenteros, J. Salomon, H. Nielsen, A. R. Johansen, NetSolP: predicting protein solubility in Escherichia coli using language models. *Bioinformatics* **38**, 941–946 (2022).

36. M. J. Buehler, MeLM, a generative pretrained language modeling framework that solves forward and inverse mechanics problems. *J Mech Phys Solids*, 105454 (2023).

37. E. Khare, C. Gonzalez-Obeso, D. L. Kaplan, M. J. Buehler, CollagenTransformer: End-to-End Transformer Model to Predict Thermal Stability of Collagen Triple Helices Using an NLP Approach. *ACS Biomater Sci Eng* **8**, 4301–4310 (2022).

38. Y. Hu, M. J. Buehler, End-to-End Protein Normal Mode Frequency Predictions Using Language and Graph Models and Application to Sonification. *ACS Nano* **16**, 20656–20670 (2022).

39. K. Guo, M. J. Buehler, Rapid prediction of protein natural frequencies using graph neural networks. *Digital Discovery* **1**, 277–285 (2022).

40. F. Y. C. Liu, B. Ni, M. J. Buehler, PRESTO: Rapid protein mechanical strength prediction with an end-to-end deep learning model. *Extreme Mech Lett* **55**, 101803 (2022).

41. A. J. Lew, M. J. Buehler, A deep learning augmented genetic algorithm approach to polycrystalline 2D material fracture discovery and design. *Appl Phys Rev* **8**, 041414 (2021).

42. E. Khare, C. H. Yu, C. Gonzalez Obeso, M. Milazzo, D. L. Kaplan, M. J. Buehler, Discovering design principles of collagen molecular stability using a genetic algorithm, deep learning, and experimental validation. *Proc Natl Acad Sci U S A* **119**, e2209524119 (2022).

43. G. Dong, G. Liao, H. Liu, G. Kuang, A Review of the Autoencoder and Its Variants: A Comparative Perspective from Target Recognition in Synthetic-Aperture Radar Images. *IEEE Geosci Remote Sens Mag* **6**, 44–68 (2018).

44. I. Goodfellow, J. Pouget-Abadie, M. Mirza, B. Xu, D. Warde-Farley, S. Ozair, A. Courville, Y. Bengio, Generative adversarial networks. *Commun ACM* **63**, 139–144 (2020).

45. J. Ho, A. Jain, P. Abbeel, Denoising Diffusion Probabilistic Models. *Adv Neural Inf Process Syst* **33**, 6840–6851 (2020).

46. G. Marcus, E. Davis, S. Aaronson, A very preliminary analysis of DALL-E 2. doi: 10.48550/arxiv.2204.13807 (2022).

47. C. Saharia, W. Chan, S. Saxena, L. Li, J. Whang, E. Denton, S. K. S. Ghasemipour, B. K. Ayan, S. S. Mahdavi, R. G. Lopes, T. Salimans, J. Ho, D. J. Fleet, M. Norouzi, Photorealistic Text-to-Image Diffusion Models with Deep Language Understanding. doi: 10.48550/arxiv.2205.11487 (2022).

48. R. Rombach, A. Blattmann, D. Lorenz, P. Esser, B. Ommer, High-Resolution Image Synthesis with Latent Diffusion Models. *Proceedings of the IEEE Computer Society Conference on Computer Vision and Pattern Recognition* **2022-June**, 10674–10685 (2021).

49. M. Z. Makoś, N. Verma, E. C. Larson, M. Freindorf, E. Kraka, Generative adversarial networks for transition state geometry prediction. *J Chem Phys* **155**, 024116 (2021).

50. T. Lebese, B. Mellado, X. Ruan, The use of Generative Adversarial Networks to characterise new physics in multi-lepton final states at the LHC. *International Journal of Modern Physics A*, doi: 10.48550/arxiv.2105.14933 (2021).

51. Z. Yang, C. H. Yu, K. Guo, M. J. Buehler, End-to-end deep learning method to predict complete strain and stress tensors for complex hierarchical composite microstructures. *J Mech Phys Solids* **154**, 104506 (2021).

52. Z. Yang, C. H. Yu, M. J. Buehler, Deep learning model to predict complex stress and strain fields in hierarchical composites. *Sci Adv* **7** (2021).

53. M. J. Buehler, FieldPerceiver: Domain agnostic transformer model to predict multiscale physical fields and nonlinear material properties through neural ologs. *Materials Today* **57**, 9–25 (2022).

54. B. Ni, H. Gao, A deep learning approach to the inverse problem of modulus identification in elasticity. *MRS Bull* **46**, 19–25 (2021).

55. B. Ni, D. L. Kaplan, M. J. Buehler, Generative design of de novo proteins based on secondary-structure constraints using an attention-based diffusion model. *Chem* **9**, 1828–1849 (2023).

56. Z. Lin, T. S. Fair, Y. Lecun, A. Rives, Deep generative models create new and diverse protein structures.

57. N. Anand, T. Achim, Protein Structure and Sequence Generation with Equivariant Denoising Diffusion Probabilistic Models. doi: 10.48550/arxiv.2205.15019 (2022).

58. B. L. Trippe, J. Yim, D. Tischer, D. Baker, T. Broderick, R. Barzilay, T. Jaakkola, Diffusion probabilistic modeling of protein backbones in 3D for the motif-scaffolding problem. doi: 10.48550/arxiv.2206.04119 (2022).

59. J. L. Watson, D. Juergens, N. R. Bennett, B. L. Trippe, J. Yim, H. E. Eisenach, W. Ahern, A. J. Borst, R. J. Ragotte, L. F. Milles, B. I. M. Wicky, N. Hanikel, S. J. Pellock, A. Courbet, W. Sheffler, J. Wang, P. Venkatesh, I. Sappington, S. V. Torres, A. Lauko, V. De Bortoli, E. Mathieu, S. Ovchinnikov, R. Barzilay, T. S. Jaakkola, F. DiMaio, M. Baek, D. Baker, De novo design of protein structure and function with RFdiffusion. *Nature 2023 620:7976* **620**, 1089–1100 (2023).

60. A. Madani, B. Krause, E. R. Greene, S. Subramanian, B. P. Mohr, J. M. Holton, J. L. Olmos, C. Xiong, Z. Z. Sun, R. Socher, J. S. Fraser, N. Naik, Large language models generate functional protein sequences across diverse families. *Nature Biotechnology 2023 41:8* **41**, 1099–1106 (2023).

61. M. Mora, A. Stannard, S. Garcia-Manyes, The nanomechanics of individual proteins. *Chem Soc Rev* **49**, 6816–6832 (2020).

62. J. Alegre-Cebollada, Protein nanomechanics in biological context. *Biophysical Reviews 2021 13:4* **13**, 435–454 (2021).

63. M. J. Buehler, S. Keten, T. Ackbarow, Theoretical and computational hierarchical nanomechanics of protein materials: Deformation and fracture. *Prog Mater Sci* **53**, 1101–1241 (2008).

64. H. Lu, B. Isralewitz, A. Krammer, V. Vogel, K. Schulten, Unfolding of titin immunoglobulin domains by steered molecular dynamics simulation. *Biophys J* **75**, 662 (1998).

65. K. C. Neuman, A. Nagy, Single-molecule force spectroscopy: optical tweezers, magnetic tweezers and atomic force microscopy. *Nature Methods 2008 5:6* **5**, 491–505 (2008).

66. E. M. Puchner, H. E. Gaub, Force and function: probing proteins with AFM-based force spectroscopy. *Curr Opin Struct Biol* **19**, 605–614 (2009).

67. M. L. Hughes, L. Dougan, The physics of pulling polyproteins: a review of single molecule force spectroscopy using the AFM to study protein unfolding. *Reports on Progress in Physics* **79**, 076601 (2016).

68. X. Zhang, L. Ma, Y. Zhang, High-Resolution Optical Tweezers for Single-Molecule Manipulation. *Yale J Biol Med* **86**, 367 (2013).

69. D. B. Ritchie, M. T. Woodside, Probing the structural dynamics of proteins and nucleic acids with optical tweezers. *Curr Opin Struct Biol* **34**, 43–51 (2015).

70. R. Tapia-Rojo, E. C. Eckels, J. M. Fernández, Ephemeral states in protein folding under force captured with a magnetic tweezers design. *Proc Natl Acad Sci U S A* **116**, 7873–7878 (2019).

71. Y.-Z. Wang, X.-M. Hou, al -, J. Valle-Orero, J. Andrés Rivas-Pardo, I. Popa, X. Zhao, X. Zeng, C. Lu, J. Yan, Studying the mechanical responses of proteins using magnetic tweezers. *Nanotechnology* **28**, 414002 (2017).

72. J. Wu, P. Li, C. Dong, H. Jiang, B. Xue, X. Gao, M. Qin, W. Wang, Bin Chen, Y. Cao, Rationally designed synthetic protein hydrogels with predictable mechanical properties. *Nature Communications 2018 9:1* **9**, 1–11 (2018).

73. M. Sikora, J. I. Sulkowska, B. S. Witkowski, M. Cieplak, BSDB: the biomolecule stretching database. *Nucleic Acids Res* **39**, D443 (2011).

74. Protein Secondary Structure - 2022. https://www.kaggle.com/datasets/kirkdco/protein-secondary-structure-2022.

75. Structural Protein Sequences. https://www.kaggle.com/datasets/shahir/protein-data-set.

76. A. Vaswani, N. Shazeer, N. Parmar, J. Uszkoreit, L. Jones, A. N. Gomez, Ł. Kaiser, I. Polosukhin, Attention Is All You Need. *Adv Neural Inf Process Syst* **2017-December**, 5999–6009 (2017).

77. RCSB PDB: Homepage. https://www.rcsb.org/.

78. P. B. Dennis, E. L. Onderko, J. M. Slocik, L. J. Bird, D. A. Phillips, W. J. Crookes-Goodson, S. M. Glaven, Proteins for bioinspired optical and electronic materials. *MRS Bull* **45**, 1027–1033 (2020).

79. T. Wang, D. He, H. Yao, X. Guo, B. Sun, G. Wang, T. Wang, D. He, H. Yao, X. Guo, B. Sun, G. Wang, Development of Proteins for High-Performance Energy Storage Devices: Opportunities, Challenges, and Strategies. *Adv Energy Mater* **12**, 2202568 (2022).

80. G. Qin, P. B. Dennis, Y. Zhang, X. Hu, J. E. Bressner, Z. Sun, W. J. Crookes-Goodson, R. R. Naik, F. G. Omenetto, D. L. Kaplan, Recombinant reflectin-based optical materials. *J Polym Sci B Polym Phys* **51**, 254–264 (2013).

81. J. Ren, Y. Wang, Y. Yao, Y. Wang, X. Fei, P. Qi, S. Lin, D. L. Kaplan, M. J. Buehler, S. Ling, Biological Material Interfaces as Inspiration for Mechanical and Optical Material Designs. *Chem Rev* **119**, 12279–12336 (2019).

82. K. Vanommeslaeghe, E. Hatcher, C. Acharya, S. Kundu, S. Zhong, J. Shim, E. Darian, O. Guvench, P. Lopes, I. Vorobyov, A. D. Mackerell, CHARMM General Force Field (CGenFF): A force field for drug-like molecules compatible with the CHARMM all-atom additive biological force fields. *J Comput Chem* **31**, 671 (2010).

83. A. V. Onufriev, D. A. Case, Generalized Born Implicit Solvent Models for Biomolecules. *Annu Rev Biophys* **48**, 275 (2019).

84. S. R. K. Ainavarapu, J. Brujić, H. H. Huang, A. P. Wiita, H. Lu, L. Li, K. A. Walther, M. Carrion-Vazquez, H. Li, J. M. Fernandez, Contour length and refolding rate of a small protein controlled by engineered disulfide bonds. *Biophys J* **92**, 225–233 (2007).

85. UniRef | UniProt help | UniProt. https://www.uniprot.org/help/uniref.

86. Z. Lin, H. Akin, R. Rao, B. Hie, Z. Zhu, W. Lu, A. Dos, S. Costa, M. Fazel-Zarandi, T. Sercu, S. Candido, A. Rives, M. Ai, Language models of protein sequences at the scale of evolution enable accurate structure prediction. *bioRxiv*, 2022.07.20.500902 (2022).

87. facebookresearch/esm: Evolutionary Scale Modeling (esm): Pretrained language models for proteins. https://github.com/facebookresearch/esm.

88. S. F. Altschul, W. Gish, W. Miller, E. W. Myers, D. J. Lipman, Basic local alignment search tool. *J Mol Biol* **215**, 403–410 (1990).

89. A. Ramesh, P. Dhariwal, A. Nichol, C. Chu, M. Chen, Hierarchical Text-Conditional Image Generation with CLIP Latents. (2022).

90. H. Zou, Z. M. Kim, D. Kang, A Survey of Diffusion Models in Natural Language Processing. (2023).

91. X. Lisa Li, J. Thickstun, I. Gulrajani, P. Liang, T. B. Hashimoto, Diffusion-LM Improves Controllable Text Generation. [Preprint] (2022). https://github.com/XiangLi1999/Diffusion-LM.git.

92. Z. Gao, C. Tan, S. Z. Li, DiffSDS: A language diffusion model for protein backbone inpainting under geometric conditions and constraints. (2023).

93. J. Dauparas, I. Anishchenko, N. Bennett, H. Bai, R. J. Ragotte, L. F. Milles, B. I. M. Wicky, A. Courbet, R. J. de Haas, N. Bethel, P. J. Y. Leung, T. F. Huddy, S. Pellock, D. Tischer, F. Chan, B. Koepnick, H. Nguyen, A. Kang, B. Sankaran, A. K. Bera, N. P. King, D. Baker, Robust deep learning–based protein sequence design using ProteinMPNN. *Science (1979)* **378**, 49–56 (2022).

94. Z. Du, H. Su, W. Wang, L. Ye, H. Wei, Z. Peng, I. Anishchenko, D. Baker, J. Yang, The trRosetta server for fast and accurate protein structure prediction. *Nature Protocols 2021 16:12* **16**, 5634–5651 (2021).

95. W. Humphrey, A. Dalke, K. Schulten, VMD: Visual molecular dynamics. *J Mol Graph* **14**, 33–38 (1996).

96. A. Paszke, S. Gross, F. Massa, A. Lerer, J. Bradbury Google, G. Chanan, T. Killeen, Z. Lin, N. Gimelshein, L. Antiga, A. Desmaison, A. K. Xamla, E. Yang, Z. Devito, M. Raison Nabla, A. Tejani, S. Chilamkurthy, Q. Ai, B. Steiner, L. F. Facebook, J. B. Facebook, S. Chintala, PyTorch: An Imperative Style, High-Performance Deep Learning Library. *Adv Neural Inf Process Syst* **32** (2019).

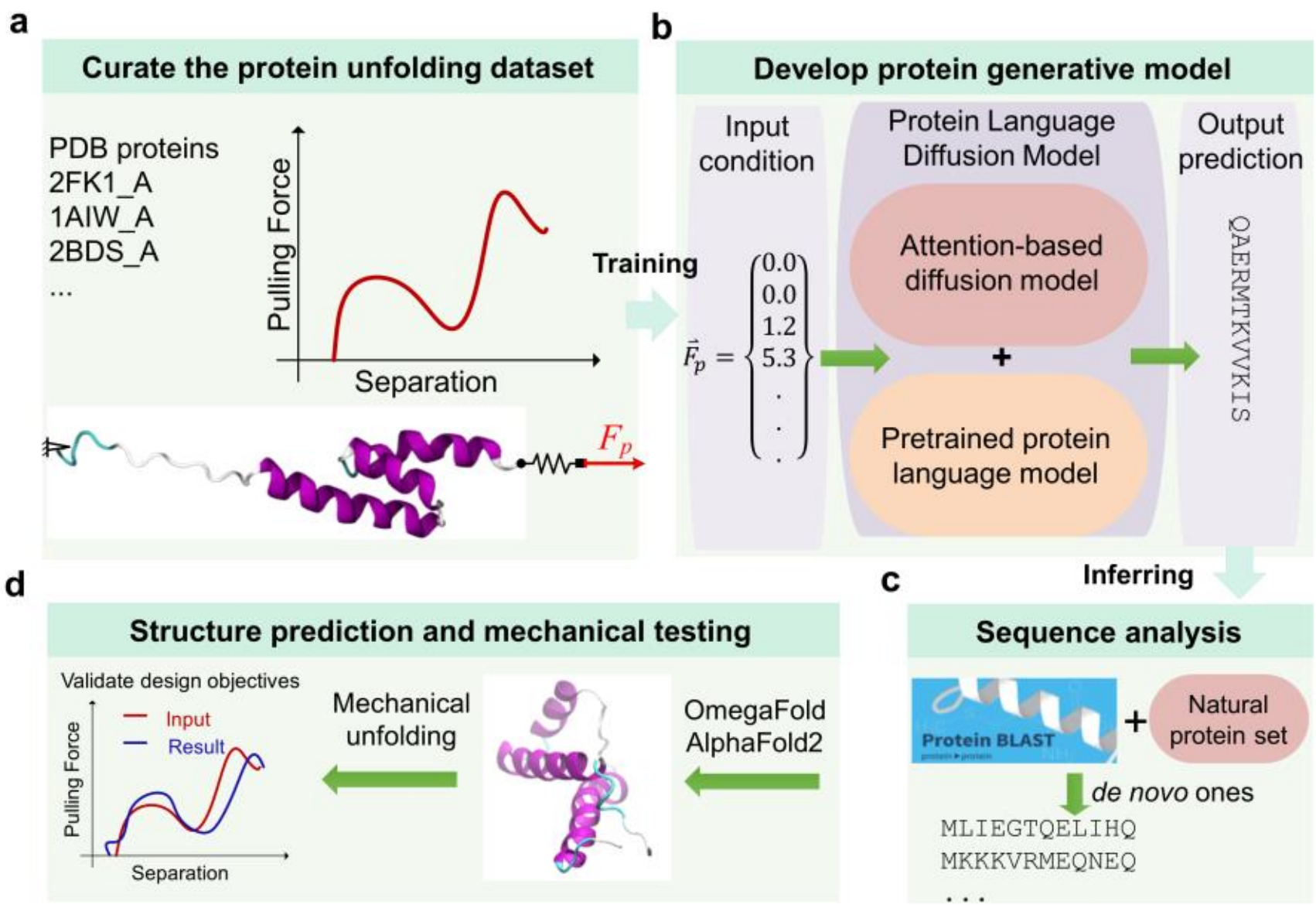

**Figure 1:** Workflow of developing the end-to-end protein generation model. **a**, curating a PDB protein dataset on their mechanical properties by unfolding protein chains by force in MD simulations. **b**, Overview of the conditioned protein language diffusion model (pLDM) developed here. **c**, analyzing the novelty of the generated protein sequences via protein-protein BLAST tests. **d**, validating the mechanical properties of the designed protein candidates using folding tools and mechanical unfolding tests.

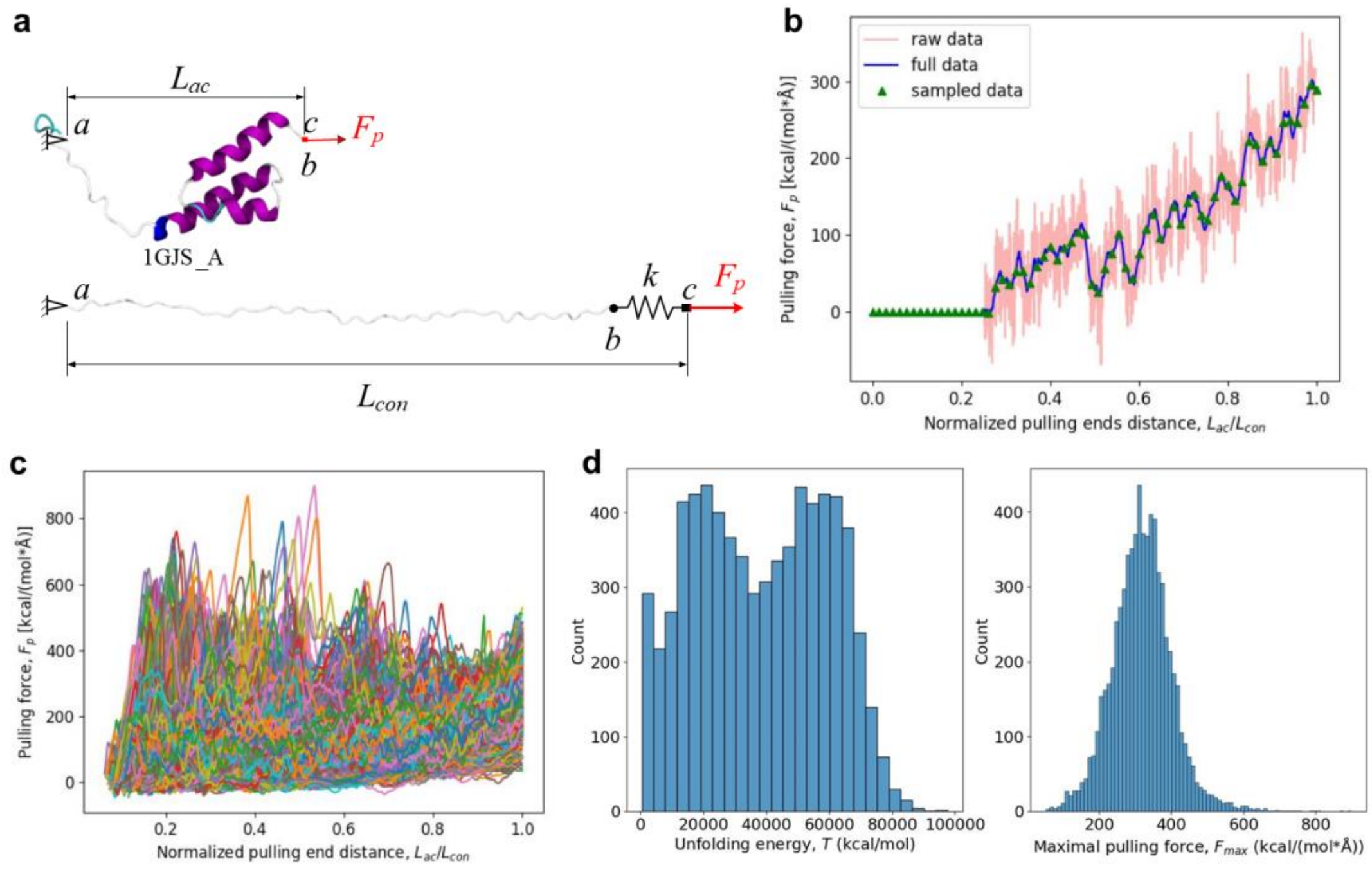

**Figure 2:** Mechanical unfolding of proteins and mechanical properties dataset curation. **a**, full-atom simulation of mechanically unfolding a PDB protein chain using steered MD. **b**, collecting (red data) and smoothing (blue data) the pulling force history during the whole unfolding process and converting it into a vector representation (green triangle dots). **c**, collecting pulling force curves during mechanically unfolding for a large member of PDB proteins. **d**, the distributions of the unfolding energy (left) and maximal pulling force (right) for the PDB protein training set developed here.

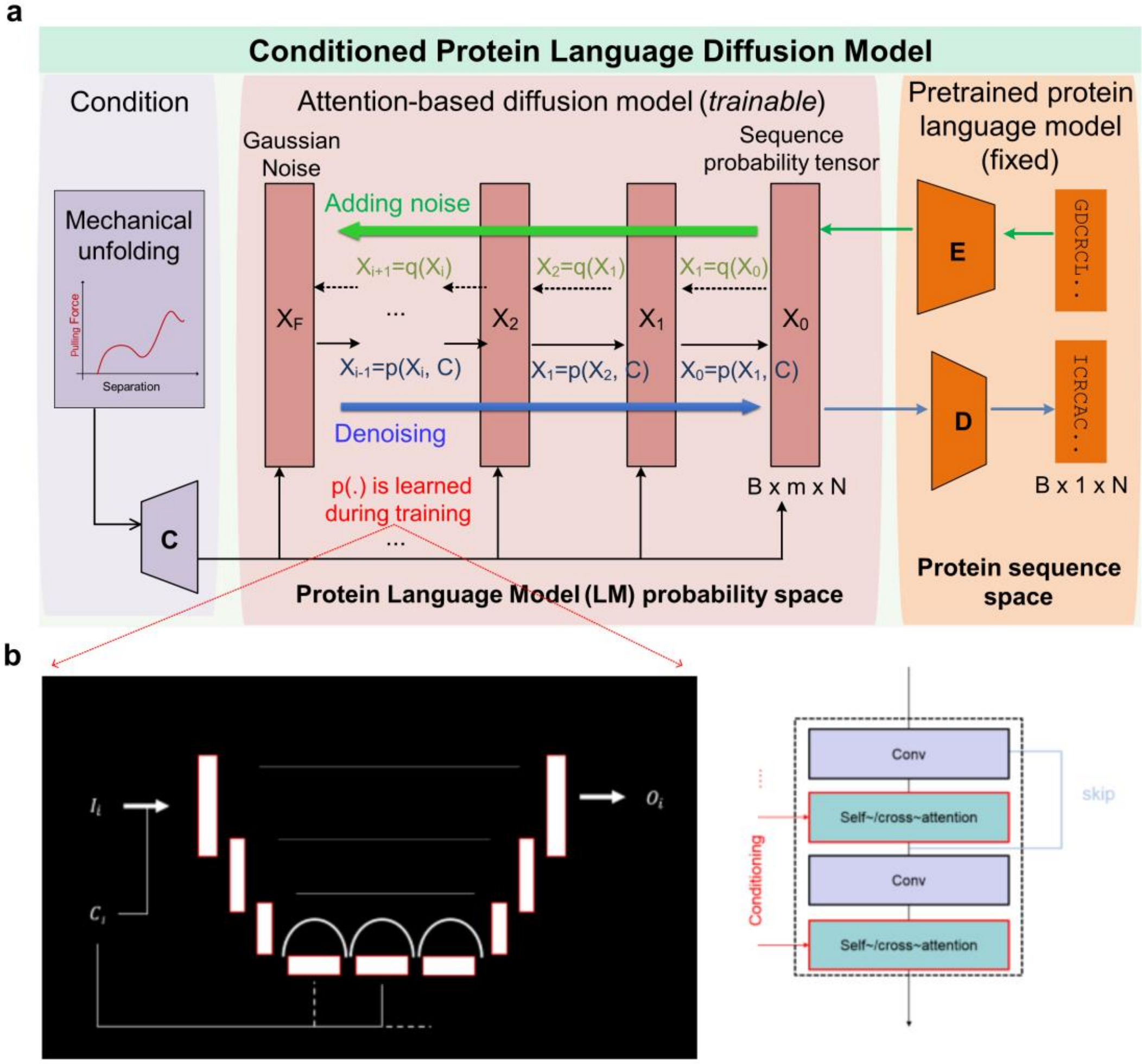

**Figure 3:** Overview of the protein language diffusion model (pLDM). a, Structure of the developed model, pLDM. It combines a protein language model (pLM) pretrained on large protein sequence data and a trainable attention-based diffusion model. We use the pretrained pLM to translate protein sequence representations between the tokenized sequence space and the word probability latent space. The diffusion model, with a building block of a 1D U-net, is trained to predict the noise added at each diffusion steps, thus gradually remove them to generate meaningful sequence representations at deployment. b, Depiction of the 1D U-net architecture that translates an input $I_i$ into an output $O_i$ under a condition set $C_i$. The model features 1D convolutional layers, as well as self-/cross-attention layers as shown on the right.

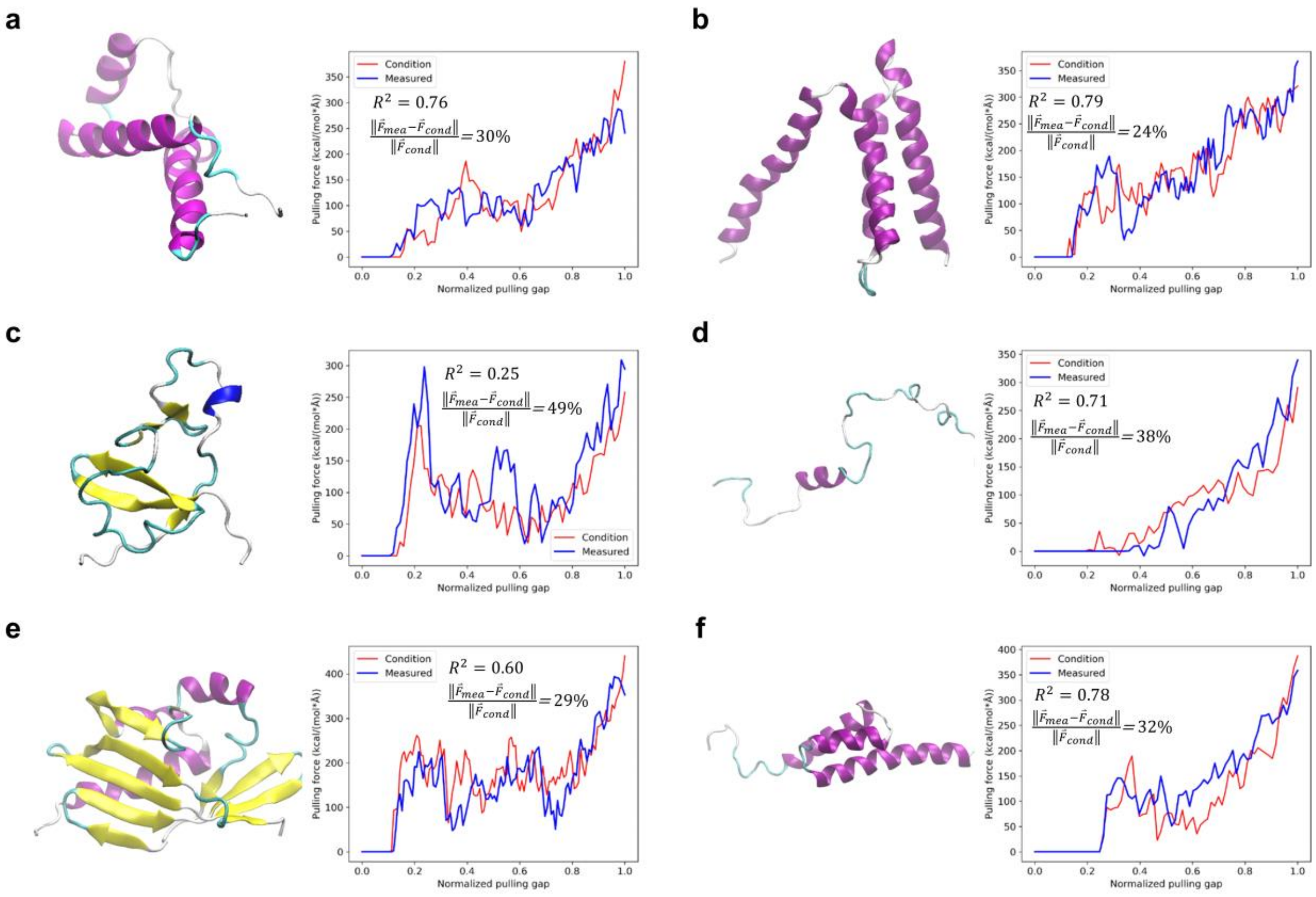

**Figure 4:** Results for protein generation based on mechanical unfolding responses of naturally existing proteins. Panels a-f show a variety of representative cases of different unfolding force paths (red curves), including the one that nearly keep increases (d), the ones that show local peaks during an overall increasing trend (a, b and f), the one that meet an oscillating plateau then increases (e) and the one reaches a high peak in the early stage (c). The proteins generated by our model demonstrate pulling force patterns (blue curves) that follow the trend of design objectives. Due to the complex and highly oscillating nature of pulling force response during mechanical unfolding of proteins, we use R2 value and relative L2 error (listed in each panel) to measure the accuracy of the design in following the overall trend and quantitative values. Corresponding to the various pulling force responses, the generated proteins show a variety of structures, including high alpha-helix content (a, b and f), a mix of beta-sheet and random coil (c), a mix of alpha-helix and random coil (d) and a mix of beta-sheet and alpha-helix (e).

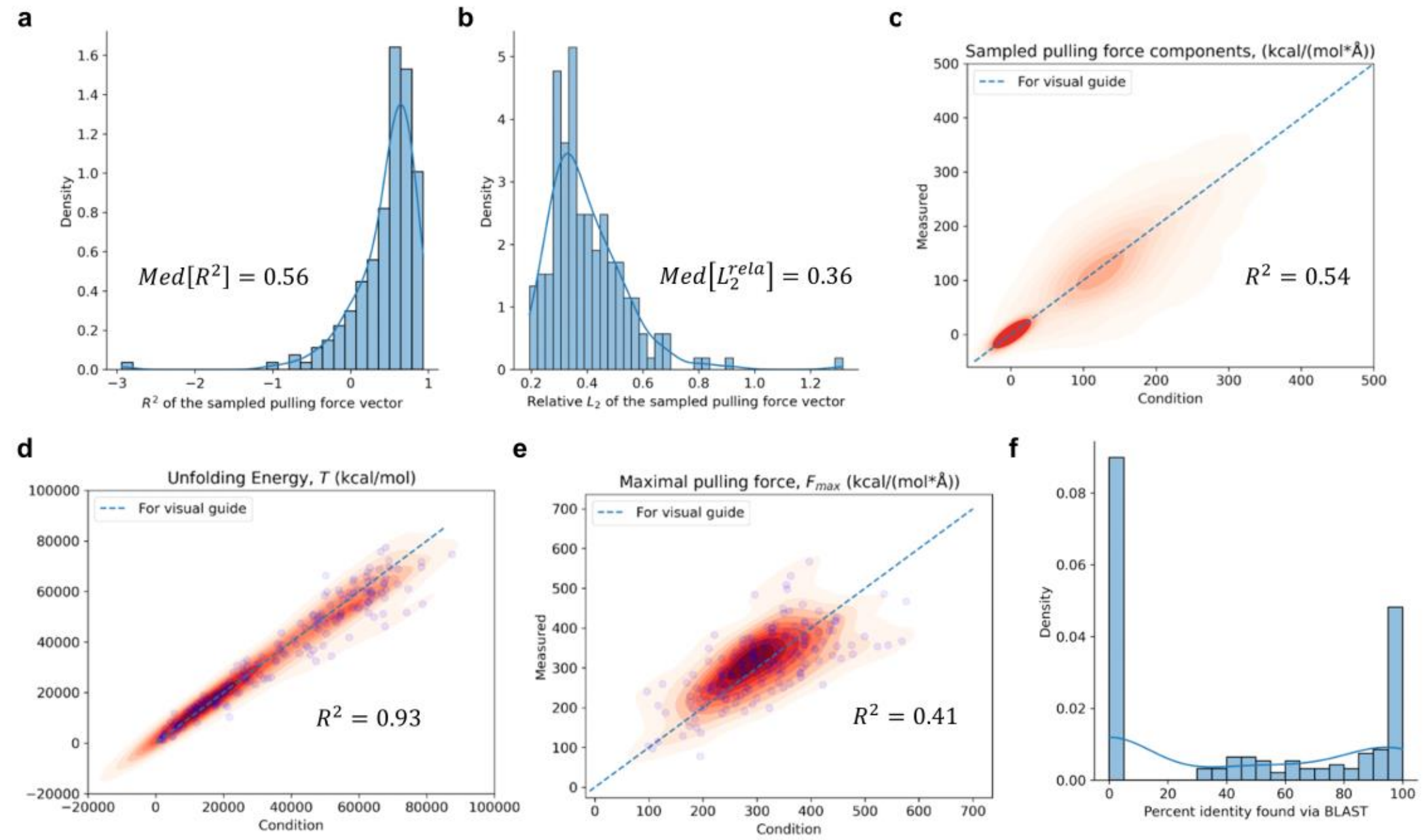

**Figure 5:** Overall quality of generating proteins based on mechanical unfolding responses that correspond to naturally existing proteins in the test set. We test the model with mechanical unfolding responses from 187 proteins in the standalone test set. On the pulling force response, **a** and **b** show the distributions of R2 (a) and relative L2 error (b) for comparing the pulling force response of each designed protein with the input condition while **c** shows the comparison in term of pulling force components for all testing cases. On the overall mechanical properties, **d** and **e** compare the designed proteins with input conditions in terms of unfolding energy (*i.e.*, toughness) and maximal pulling force (*i.e.*, strength). On the novelty of the designed sequences, **f** shows the distribution of the highest percent identity found via BLAST test.

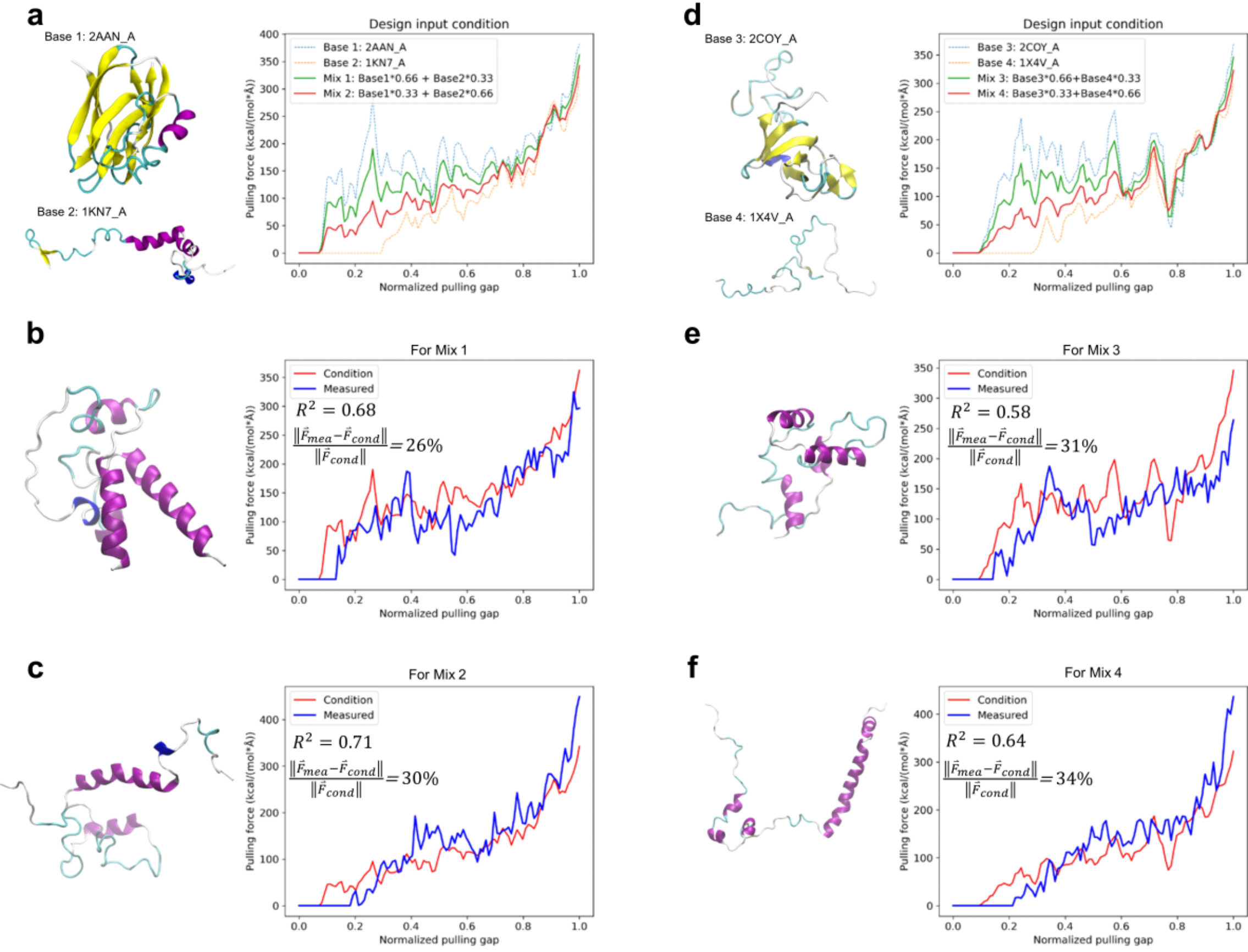

**Figure 6:** Results for protein generation based on de novo mechanical unfolding responses. For de novo design input, in **a** and **d**, we construct novel mechanical unfolding responses by mixing existing ones at a different ratio and changing the targeted sequence length. Here, we intentionally choose pulling forces of different patterns, one that monotonically increases (base 2 in a and base 4 in d) and the other that first reaches a plateau then increases (base 1 in **a** and base 3 in **d**) as the bases to mix. By choosing the mixing ratio, the de novo inputs cover a transition between these two pulling force pattern. **b-c** and **e-f** show the results of the generated proteins. On the pulling force, the generated protein follows the design input closely in terms of both the overall pattern and quantitative values throughout the unfolding process. On the structure, interestingly, the generated proteins show a mix of alpha-helix and random coil, which is different from some of those base proteins of high beta-sheet content. On the sequence length, the generated proteins have a length that is different from both base proteins.

**Table 1:** Results of the BLAST analysis for the various generated proteins (from **Figure 4**) based on existing mechanical unfolding responses. Given the pulling force vectors of existing proteins as the design condition, the model still shows high probability in predicting sequences that show little similarity to existing proteins as can be seen from the BLAST results (a-d and f). For other cases, sequences with some similarity to known proteins can be predicted (e).

| Case | Sequence | BLAST result: the sequence producing the most significant alignment | |
|---|---|---|---|
| | | among PDB proteins | beyond PDB proteins |
| a | MLIEGTQELIHQKLAKGKTVLVQ<br>RYVAKGLQVDDNTEDLLANAKNY<br>LNPDQIERSIAYAQKIEEMEGDD<br>MFKVALV | -- | None significant similarity found (NSSF) |
| b | MKKKVRMEQNEQKKQVYQELNDK<br>VENDEALAPKSVALYIAALKEKE<br>EAGKIPHHFNLLERLKLTITSCR<br>FFLLKIQNNDTKLQKRRKFIDET<br>IQLAREIYEKQDNK | -- | NSSF |
| c | MGKITPVVLAGGKQKEDEETLDG<br>GEILTKDGKTLKLISDAQVAVMN<br>VKQVQEGTYEGSQVIEEDGVRGN<br>YVSYVGK | -- | NSSF |
| d | GSSGSSGRDVTQQTNKCCRRCSR<br>KPHCCIKAWRPRSSDLYYHEKHT<br>HSGPSSG | -- | NSSF |
| e | MNTPEHMTAVVQRYVAALNGGDL<br>DGIVALFADDATVEDPVGFQNVS<br>GKAADANFYESPGFLDLVKALTG<br>PVRAFGNEKFFAMIVFFEYEGTK<br>TVVAGIDHIRFNGAGKVVSMRAY<br>FDEKNIHASA | 100% query cover, 73.6% identical with 8CH0 | -- |
| f | MPWHHHGSSGLVQTGMAATGLKD<br>FIVEAYPKKPDDIIKVCRSEPSA<br>GYWWCEDVQNEVKQKCLSKKQRQ<br>VKAQ | -- | NSSF |

**Table 2:** Results of the BLAST analysis for the various generated proteins (from **Figure 6**) based on *de novo* mechanical unfolding responses. Given the *de novo* pulling force vectors as the design condition, the model shows high probability in predicting *de novo* sequences that show little similarity to existing proteins as can be seen from the BLAST results (For Mix 1-4).

| Case | | BLAST result: the sequence producing the most significant alignment | |
|---|---|---|---|
| | | among PDB proteins | beyond PDB proteins |
| For Mix 1 | MDDALLLKAMQQLLLAPIRVKED<br>DPLVRRDAAIGFAPDGVRVDFEY<br>TAKVDLAKATLDEVGLKGANTTQ<br>PPRPIANKLPPPIVVLASKLLEI<br>YKELKQL | -- | None significant similarity found (NSSF) |
| For Mix 2 | GSSGSSGYGRQVTTRSPRETTSL<br>SFDIDREPMEFNQLKAQELMMPF<br>NLKALDTGRFNRPLQFVEQAKGK<br>MEKALLKKATDPVQALPKKDLSE<br>GISPKMG | -- | NSSF |
| For Mix 3 | MMDTPKLMDELKDYAPQPARRAL<br>NLTNPRTAAVPEKTGDDVAPFFD<br>HAAEKENLGFHHEVANDNWESEA<br>KFLKLTKVPVSPQVIYNAAGLLF<br>EAARKTP | -- | NSSF |
| For Mix 4 | GSSGSSGPSTYKNPGDRFFTTSY<br>FTDPELEAGQFEVRTKDKMLNGI<br>TLLQQKPCGKSCELFVDQNKKAV<br>EEKKKKLMLTQMAAYYQQDDLTM<br>ASGPSSG | -- | NSSF |

**Table 3:** Performance comparison between the current pLDM with one-shot prediction (last column) and the protein diffusion model with one-shot prediction (3rd last column) as well as iterative predictions (2nd last column). Tested with the existing unfolding responses from the test set, the pLDM shows an overall better performance in fulfilling the design target by achieving the best results (in red, the minimal for errors and the maximal for $R^2$) in the most rows considering mechanical properties as well as the detailed pulling force responses.

| Performance on the test set | | | | AM1: pDM with one-shot generation | AM2: iterative prediction with pDM based protein designer and predictor | The current model: pLDM with one-shot prediction |
|---|---|---|---|---|---|---|
| Toughness (*i.e.*, unfolding energy) | | $R^2$ | | 0.86 | 0.87 | 0.93 |
| | | Relative $L_1$ error | Mean | 0.151 | 0.150 | 0.147 |
| | | | Median | 0.121 | 0.123 | 0.102 |
| Strength (*i.e.*, $F_{max}$) | | $R^2$ | | 0.09 | 0.17 | 0.41 |
| | | Relative $L_1$ error | Mean | 0.188 | 0.164 | 0.188 |
| | | | Median | 0.151 | 0.113 | 0.149 |
| Pulling force vectors | As vectors | $R^2$ | Mean | 0.427 | 0.418 | 0.452 |
| | | | Median | 0.526 | 0.522 | 0.563 |
| | | Relative $L_2$ error | Mean | 0.399 | 0.402 | 0.398 |
| | | | Median | 0.377 | 0.382 | 0.362 |
| | As components | $R^2$ | | 0.476 | 0.499 | 0.537 |

**Table 4:** A short summary of the performance of protein language diffusion model developed in the present work and other models discussed.

| Model name | Tested input conditions | Design accuracy | Design novelty |
|---|---|---|---|
| The developed model: protein language diffusion model using one-shot design | Mechanical unfolding responses from naturally existing proteins | Good agreement with the designed pulling force responses as well as the strength and toughness in trend and values | Tend to generate *de novo* ones, but can also rediscover ones that show some similarity to existing proteins |
| | *De novo* mechanical unfolding responses | Similar to the above | More probable to discover *de novo* sequences |
| AM1: Protein diffusion model using one-shot design | Mechanical unfolding responses from existing proteins | Slightly weaker than AM2 | -- |
| AM2: Protein diffusion model using multi-shot iterative design | Mechanical unfolding responses from existing proteins | Weaker than the developed model | -- |

# ForceGen: End-to-end *de novo* protein generation based on nonlinear mechanical unfolding responses using a protein language diffusion model

Bo Ni[1], David L. Kaplan[2], Markus J. Buehler[1,3]*

[1] Laboratory for Atomistic and Molecular Mechanics (LAMM), Massachusetts Institute of Technology, 77 Massachusetts Ave., Cambridge, MA 02139, USA

[2] Department of Biomedical Engineering, Tufts University, Medford, MA 02155, USA

[3] Center for Computational Science and Engineering, Schwarzman College of Computing, Massachusetts Institute of Technology, 77 Massachusetts Ave., Cambridge, MA 02139, USA

*mbuehler@MIT.EDU

# SUPPLEMENTARY INFORMATION

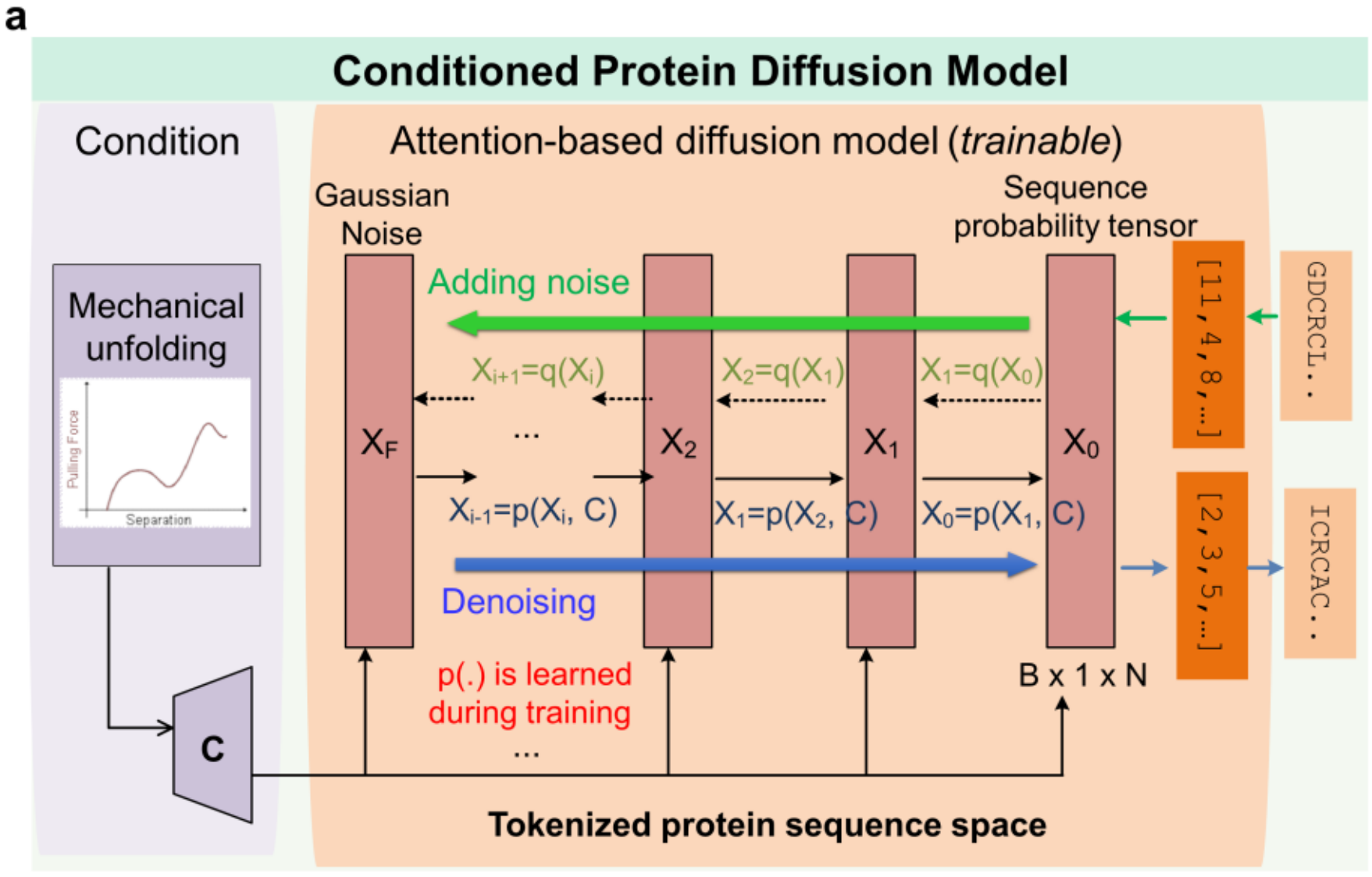

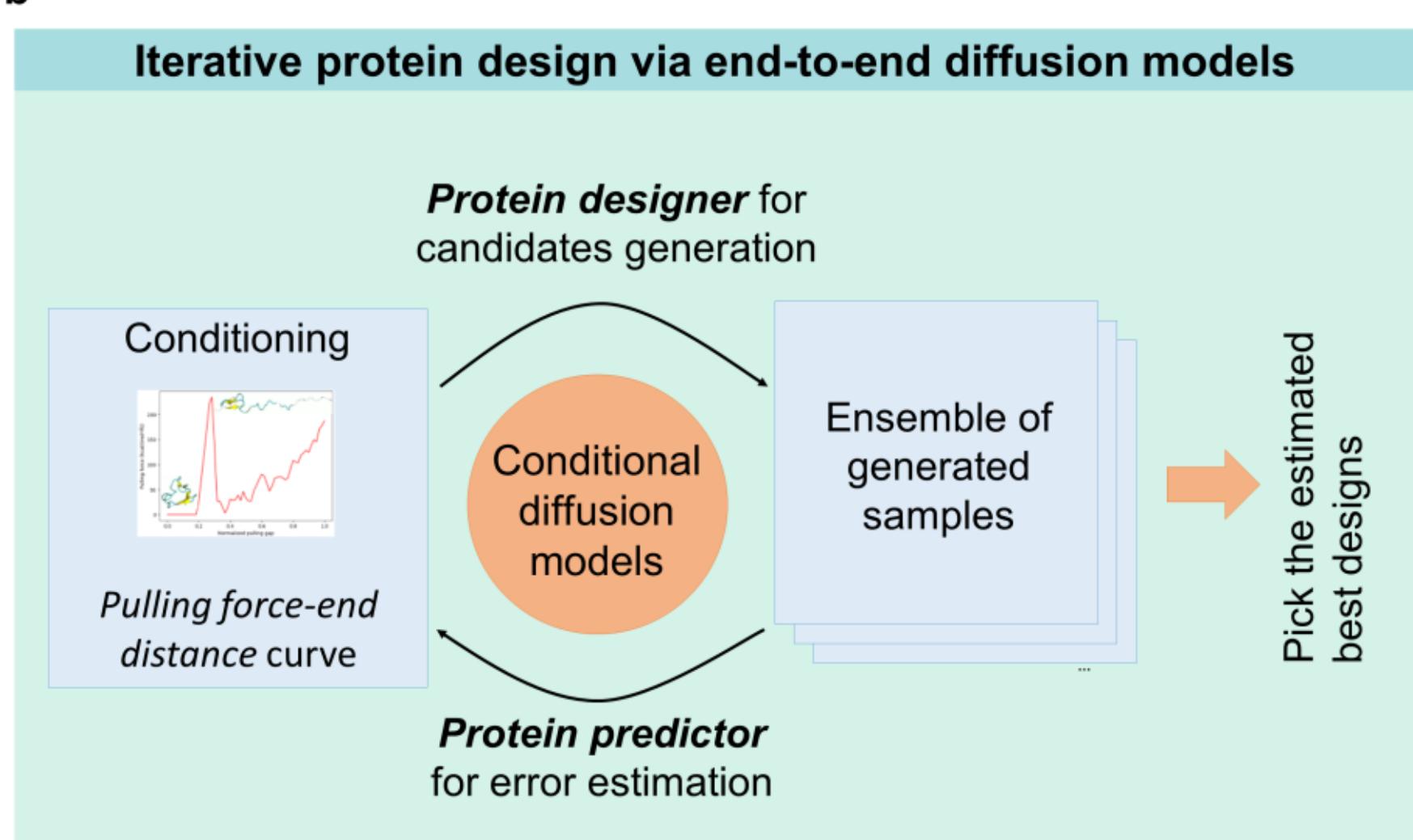

**Figure S1**: Overview of adjusting protein diffusion models for protein design based on mechanical unfolding response design objectives. a, Structure of the adopted protein diffusion model. b, An iterative design scheme using protein diffusion model based protein designer (*i.e.*, designing protein sequences for given pulling force vectors) and protein predictor (*i.e.*, predicting pulling force vector based on given protein sequences).